\documentclass[10pt]{article}
\usepackage{graphicx}
\usepackage{amsmath}
\usepackage{amsfonts}
\usepackage{amssymb}
\usepackage{cite}

\begin{document}
\title{Osmotic force resisting chain insertion in a colloidal suspension}
\author{M. Castelnovo, R. K. Bowles\thanks{present address: Courant Institute of Mathematical Sciences, NYU, New York, New York 10012.}, H. Reiss and W. M. Gelbart \\
\textit{Department of Chemistry and
Biochemistry},\\ 
\textit{University of California Los Angeles,}\\ \textit{Los Angeles, California
90095}}
\maketitle
\begin{abstract}
We consider the problem of inserting a stiff chain into a colloidal suspension of particles that interact with it through excluded volume forces.  The free energy of insertion is associated with the work of creating a cavity devoid of colloid and sufficiently large to accommodate the chain. The corresponding work per unit length is the force that resists the entry of the chain into the colloidal suspension. In the case of a hard sphere fluid, this work can be calculated straightforwardly within the scaled particle theory; for solutions of flexible polymers, on the other hand, we employ simple scaling arguments. The forces computed in these ways are shown, for nanometer chain and colloid diameters, to be of the order of tens of $pN$ for solution volume fractions of a few tenths. These magnitudes are argued to be important for biophysical processes such as the ejection of DNA from viral capsids into the cell cytoplasm. 
\end{abstract}
\section{Introduction}
DNA is known to be packaged at very high concentrations in most viruses \cite{lodish}. In the particular case of bacterial viruses, or bacteriophages ("phage"), the high free energy density of double-stranded DNA bent and crowded inside the viral capsid is thought to be responsible for the initial injection step of the viral genome into its host cell.  In these instances the virus binds to its specific receptor on the outer cell membrane, its capsid is opened, and the stored energy of confinement drives the ejection process. As the ejection proceeds the driving force decreases because of the progressive relief of stress inside the capsid.  Later steps in the translocation of the full viral genome from the capsid into the cell vary from one type of phage to another.  For example, it has been shown in the case of T7 bacteriophage that the translocation is coupled to the transcription of viral DNA by the host cell machinery, which exerts a force pulling the DNA inside the cell \cite{garcia}.  In other cases, such as T5, the translocation can occur without the intervention of an active process, e.g., via interaction with the DNA-binding proteins present at high concentrations in the cell cytoplasm \cite{labedan}.

Over the last couple of decades there have been several attempts to estimate theoretically the energy cost of confining DNA inside a viral capsid \cite{bloomfield,grosberg,odijk}. Recent experiments \cite{smith} and simulations and phenomenological theory \cite{kindt,shelly,phillips} on the packaging of viral DNA indicate that the stress associated with fully loaded capsids corresponds to ejection forces on the order of tens of $pN$ and to internal pressures of several tens of $atm$.  Osmotic pressures of this magnitude are commonly realized in polymer solutions at moderately high concentration, thereby raising the following simple question \cite{raspaud}: is it possible to stop the DNA ejection from a viral capsid by "opposing" the internal pressure with that exerted by a colloidal suspension? This question can be formulated more precisely in terms of the associated \textit{forces}: can a colloidal suspension (hard spheres or flexible polymers) produce a force opposing that driving the DNA out of the capsid? This problem is also relevant for \textit{in vivo} phenomena, since the cell cytoplasm, containing high concentrations of various components (mainly proteins), has been modeled as a colloidal suspension to evaluate the effect of the ``macromolecular crowding'' on various biochemical reactions \cite{ellis}.

Our aim in this work is to formulate a theory for the force resisting chain insertion into a colloidal suspension. This force is evaluated by calculating the free energy of insertion of the chain in the suspension, more explicitly the work required to "grow" the particle to its final size and shape in the solution \cite{landau}. This concept has long played a fundamental role in the evolution of modern theories of simple liquids \cite{henderson}. For the case of hard sphere fluids, one of the most physically appealing and natural tools for calculating this work of insertion is the Scaled Particle Theory (SPT) \cite{howard}. The force resisting the injection is then given by the derivative of this work with respect to inserted chain length. If the colloidal suspension is a solution of flexible polymers, the work of insertion is evaluated naturally through scaling laws. As will be shown in this paper, the force resisting chain injection in both of these instances can indeed be comparable to the force driving DNA out of viral capsids \cite{kindt,shelly,phillips}.  This result accounts for recent experiments in which DNA ejection from bacteriophage ($\lambda$) is shown to be suppressed by a sufficiently high concentration of flexible polymers (polyethyleneglycol - PEG) in solution \cite{alex}. In these measurements an enzyme is also present which digests the ejected DNA into free nucleotides, so that one does not have to deal with any of the complications of PEG-induced DNA condensation \cite{khokhlov}, since only short (sub-persistence length), isolated, pieces of DNA are involved. This also validates our modeling of the ejected DNA as a stiff chain, in calculating the force resisting its entry into the colloidal suspension. More systematic analyses of these experiments will be provided elsewhere.  Here we focus on the relationship between resisting force and osmotic pressure, comparing and contrasting the situations of hard sphere versus flexible polymer solutions.

Stiff chains -- rods -- in both hard sphere suspensions and flexible polymer solutions have been treated previously, using SPT and scaling theory, respectively.   But the aims have invariably been the calculation of \textit{phase behavior} -- say, demixing via depletion interactions in rod/sphere mixtures \cite{lek1} or chain condensation via effective attractions in solutions of flexible polymers \cite{devries}.  Our aim, on the other hand, is to describe directly the \textit{forces} that arise in these colloidal systems, acting directly to resist the entry of chains.  The paper is organized as follows. In the next section the general principle of the SPT is outlined briefly and results from earlier work are quoted for the calculation of the work of insertion of a spherical cavity in a hard sphere fluid. In section \ref{spc} we generalize this approach to the case of a cavity whose shape is spherocylindrical, calculating the force resisting entry of the chain from the insertion energy per unit length. The work of inserting a cylinder into a flexible polymer solution is then estimated from simple scaling arguments in section \ref{workpol}, again generalizing from the spherical solute case. The forces derived in both instances -- hard sphere and flexible polymer solutions -- are discussed in the final section, as a function of chain diameter, colloid size, and concentration.  Numerical estimates are made for the case of DNA and PEG solutions, before we conclude with some remarks about connections to other biophysical phenomena involving the translocation of stiff chains into concentrated colloidal suspensions.
\section{Work of insertion of a spherical cavity in a hard sphere fluid}
We present in this section the SPT calculation of the work of insertion of a spherical cavity of arbitrary radius $r$ in a hard sphere fluid of concentration $c_P$, the diameter of the spheres being $a$. 
For the sake of simplicity, we choose $kT$ as the energy unit throughout this paper. The cavity is defined by a spherical volume $\frac{4\pi r^3}{3}$ free of any centers of hard spheres. For a small enough cavity, $r<a/2$, the work of insertion is related exactly to the free volume fraction by
\begin{equation}
\label{worksmall}
W_0(r)=-\ln \left(1-c_P \frac{4\pi r^3}{3}\right), \, \, r<a/2
\end{equation}
since there is only room for zero or one sphere. For bigger cavities, more spheres can be accomodated, so that exact determination of $W(r)$ would require the knowledge of $n$-body correlation functions \cite{howard}.

One of the key achievements of the SPT is the providing of a physically motivated approximation for the work of insertion without determining the whole structure of the hard sphere fluid.
Within the SPT it can be shown that the form of the work of insertion of a spherical cavity is restricted by a certain number of exact conditions. By choosing a simple form for the work of insertion satisfying those exact conditions, one expects to have an accurate approximation for this work.
Since we want to calculate the work of insertion for an arbitrary value of cavity radius, it is judicious to choose the asymptotic value for $r\rightarrow \infty$ given by the sum of a pressure-volume work and a surface tension contribution:
\begin{equation}
\label{formwork}W_{\infty}(r)=pV+\sigma \left(1-\frac{2\delta}{r }\right)S
\end{equation}
The quantities $p,\sigma,\delta$ are respectively: the pressure of the hard sphere fluid; its surface tension in the limit of flat surface; and the Tolman length associated with the curvature of the surface \cite{tolman}. $V=4\pi r^3/3$ and $S=4\pi r^2$ are the volume and surface of the cavity. As shown below, by using three conditions derived exactly through the SPT, one can determine the hard sphere fluid properties. 

The first two conditions are the continuity of the work's first two derivatives at $r=a/2$. Therefore, by matching the macroscopic values of these quantities -- see Eq.\ref{formwork} -- with their exact microscopic values for $r\le a/2$ -- see Eq.\ref{worksmall}, we obtain
\begin{equation}
\label{sig}\sigma =-\, \frac{9y^2(1+y)}{2\pi a^2 (1-y)^3} 
\end{equation}
and
\begin{equation}
\label{del}\delta=\frac{ya}{2(1+y)}
\end{equation}
$y=\pi a^3c_P/6$ being the volume fraction of spheres.
Note that the surface tension of a hard sphere fluid is negative, reflecting the absence of attractive interactions holding the particles together as in a ``real'', cohesive, fluid. The pressure is found by another exact relation, 
\begin{equation}
\frac{p}{c_P}=1+4y\,g(a),
\label{pressure}
\end{equation}
where $g(r)$ is the pair correlation function, proportional to the probability of finding the center of a hard sphere at a distance $r$ from the origin, provided that there is already another particle at the origin \cite{henderson}. $g(a)$ can in turn be expressed, without approximation, in terms of the first derivative of the work of insertion:
\begin{equation}
\label{gr}g(a)=\frac{W'(a)}{4\pi a^2c_P}
\end{equation}
In this way one obtains the SPT equation of state for the hard sphere fluid
\begin{equation}
\label{py}
p =\frac{6}{\pi a^3}\, \frac{y+y^2+y^3}{(1-y)^3}
\end{equation}
The second and third virial coefficients derived from the expansion of Eq.\ref{py} for low $y$ are known to be exact, while up to the fifth order the discrepancy  is less than 5 \%. Therefore as long as the volume fraction of particles is not too high, the SPT equation of state provides an accurate description of the hard sphere fluid. Eq. \ref{py} is also known to be identical to the pressure equation of state obtained via the Percus-Yevick closure \cite{rosenfeld,percus,wertheim}. 
\section{Work of insertion of a spherocylinder in a hard sphere fluid}
\label{spc}
We generalize in this section the SPT formalism to calculate the work of insertion of a spherocylindrical cavity  in a hard sphere fluid. The spherocylinder considered here has a total length $La+(1+D)a$ and a diameter $(1+D)a$. When $L,D=0$, the cavity is just a sphere of diameter $a$, so that the exact form of the work of insertion is still given by Eq.\ref{worksmall}. For bigger cavities, we can compute it approximately by using, as in the original SPT approach, exact conditions restricting its form. A quite natural choice is the work of insertion of a macroscopic spherocylinder in a hard sphere fluid
\begin{equation}
\label{defw}
W(D,L)=pV +\sigma_1 \left(1-\frac{2\delta_1}{R_1 }\right)S_1 +\sigma_2 \left(1-\frac{\delta_2}{R_2}\right)S_2
\end{equation}
We explicitly choose different coefficients associated with the spherical (indices 1) and cylindrical (indices 2) part of the cavity, in order to start with the most general form of the macroscopic work. As shown below, these quantities turn out to be equal in the present calculation. $R_1$ and $R_2$ are respectively the curvature radii of the spherical and cylindrical parts of the cavity. This form of the work of insertion again corresponds to a pressure-volume work and a surface contribution. Note also that this form involve an expansion of the work  in powers of $D$ and $L$, since $V=\pi a^3\left((1+D)^3+\frac{3}{2}L(1+D)^2\right)/6$, $S_1=\pi a^2 (1+D)^2$ and $S_2=\pi a^2 L(1+D)$ \cite{rosenfeld,lek1}. The coefficients are found by matching the first two derivatives of the work at $D,L=0$. The results read
\begin{equation}
\sigma_1  =  \sigma_2 = \frac{3y(2+y)}{2\pi a^2 (1-y)^2}-\frac{pa}{2} 
\end{equation}
and
\begin{equation}
\delta_1 = \delta_2 =\frac{a}{4}\,\frac{p\pi a^3-\frac{6y(1+2y)}{(1-y)^2}}{p\pi a^3-\frac{3y(2+y)}{(1-y)^2}}
\end{equation}
Finally, the pressure involved in the preceding equations is found by using Eqs.\ref{pressure} and \ref{gr}, leading again to expressions for $p,\sigma$ and $\delta$ given by Eqs.\ref{py}, \ref{sig} and \ref{del}. The fact that the coefficients are simply the pressure, surface tension and Tolman length of a pure hard sphere fluid indicates that, for such a simple geometry, the shape of the cavity does not really matter for the work of insertion, and the macroscopic form Eq. \ref{formwork} obtains as before. The work of insertion Eq.\ref{defw}, together with Eq.\ref{py},\ref{sig} and \ref{del}, is the same as the one derived previously for phase behavior studies of rod/sphere mixtures \cite{lek1,lek2}. The work of insertion in the present study will be used in section \ref{discussion} to evaluate the force resisting the insertion of a spherocylinder into a hard sphere fluid of given volume fraction.
\section{Work of insertion of a cylinder in a flexible polymer solution}
\label{workpol}
Unlike in the hard sphere fluid case, no exact conditions restricting the form of the actual work of insertion of a cavity of arbitrary shape or size in a polymer solution has been derived yet. Moreover it is not straightforward to try to map the SPT principle to the polymer case because the cavity created around one chain excluding other chains is itself a statistical object, \textit{i.e.} depends on the conformation of the polymer. Nevertheless, the work of insertion of a cavity of spherical or cylindrical shape can be calculated using either pure scaling arguments \cite{joanny,sear,devries} or renormalization group arguments \cite{eisenriegler1,eisenriegler2}, \textit{i.e.} specific polymer physics tools. We briefly present here scaling arguments that allow one to approximate the work of insertion of a cylindrical cavity in a polymer solution by superimposing asymptotic behaviors of the work. This means implicitly that all numerical prefactors will be set equal to unity. 

We consider first the interaction of a cylinder of length $L$ and diameter $D$ with a flexible chain of radius of gyration $R_g$ in the limit of thin cylinder $R_g>>D$. The monomer size and the degree of polymerization are, respectively, $b$ and $M$. The interaction energy is proportional to the number of contacts between the cylinder and polymer. If the two objects occupy the same average volume $R^3$, then their interaction energy within a simple mean-field approximation is given by
\begin{equation}
\frac{F_{int}}{kT}\simeq R^3 \phi_{cyl}\phi_{coil}
\end{equation}
where the quantities $\phi_{cyl}\simeq \frac{LD^2}{R^3}$ and $\phi_{coil}\simeq \frac{Mb^3}{R^3}$ are respectively the volume fraction of the cylinder and the chain. Therefore we expect this interaction energy to be linear in both the length of the cylinder and the degree of polymerization of the coil. The mean-field argument used here is justified since the interaction energy scales with the common size of the interacting objects $R$ as $F_{int}\sim kT \frac{D^2b^{4/3}}{R^{1/3}}$, and therefore decreases as the sizes of the objects are increased \cite{joanny}. Similar arguments can be used to show for example that the contribution of binary contacts in a single chain is irrelevant if the dimension of the embedding space $d$ exceeds 4, while in the opposite case simple mean field approximations fail \cite{degennes}. 

It is straigthforward to show that regardless of whether the cylinder is longer or smaller than the radius of gyration of the chain, the interaction energy must still vary linearly both with the length of the cylinder and the degree of polymerization of the chain. With these requirements, a cross virial coefficient between cylinder and polymer of general scaling form $v_{cyl-ch}\sim L^{\alpha}D^{\beta}R_g ^{\gamma}$ is uniquely determined and becomes
\begin{equation}
v_{cyl-ch}\sim LD^{1/3}R_g ^{5/3}
\end{equation}
In a dilute solution of polymers of monomer concentration $c$, the work of insertion is then given by
\begin{equation}
W\sim LD^{1/3}R_g ^{5/3}\frac{c}{M}
\end{equation}

The previous arguments can easily be generalized to the case of semi-dilute solution of polymers where chains overlap. The overlap concentration is given by $c^* b^3 =y^* \sim M^{-4/5}$ \cite{degennes}. In this limit, the polymer solution is characterized by the average mesh size of the temporary network formed by the chains, or equivalently the correlation length, that scales like $\xi \sim b(cb^3)^{-3/4}$ \cite{degennes}. A chain subunit of size $\xi$ is called a semi-dilute blob, so that the semi-dilute solution can be envisioned as a compact packing of such blobs, with the statistics of monomers within a blob given by those of an isolated chain: if $g$ is the number of monomers per blob, then $\xi \sim g^{3/5}b$. The cross virial coefficient between the cylinder and the blobs is given by $v_{cyl-bl}\sim LD^{1/3}\xi ^{5/3}$, so that the work of insertion in a semi-dilute solution reads
\begin{equation}
W\sim LD^{1/3}\xi ^{5/3}\frac{c}{g}\sim \frac{LD^{1/3}}{\xi^{4/3}}
\end{equation}

As noted by de Vries, the result for the work of insertion in the dilute and in the semi-dilute case are the same:
\begin{equation}
W\sim \frac{LD^{1/3}}{b^{4/3}}y
\end{equation}
with $y=cb^3$ \cite{devries}.
This work is independent of the chain degree of polymerization, and is linear in the volume fraction of monomers. This follows from the fact that the insertion of a cylinder is dominated by the depletion layer around it of order $D$ and is a local process, as long as its diameter $D$ is smaller than the correlation length of the polymer solution ($R_g$ in dilute solution and $\xi$ in semi-dilute solution). 

When $R_g,\xi<<D$ on the other hand, the depletion layer around the cylinder is of order the correlation length of the polymer solution, so that the work of insertion is dominated by the pressure-volume work and surface tension contributions. As in the case of a large cavity in the hard sphere fluid, this corresponds to the macroscopic value of the work. Therefore this work is written
\begin{equation}
W\sim \left\{ 
\begin{array}{ll}
\frac{c}{M}LD^2+\sigma_d LD \, \, \mathrm{if} \, \, c<<c^*\\
\frac{1}{\xi^{3}}LD^2+\sigma_{sd}LD \, \, \mathrm{if} \, \, c>>c^*
\end{array}
\right.
\end{equation}
The surface tensions scale like $\sigma_d \sim \frac{c}{M}R_g$ and $\sigma_{sd}\sim \xi^{-2}$ \cite{eisenriegler2}.
Therefore the different asymptotic behaviors of the work of insertion of a cylinder in a polymer solution can be summarized as follows.
\begin{itemize}
\item if $R_g<D$,
\begin{equation}
W_{rod}= \left\{ 
\begin{array}{ll}
\frac{y}{M}\frac{LD^2}{b^3} +\frac{yR_g}{Mb^3}DL\, \, \mathrm{for} \, \, y<y^* \\
y^{9/4}\frac{LD^2}{b^3} +y^{6/4}\frac{DL}{b^2}\, \, \mathrm{for} \, \, y>y^*
\end{array}
\right.
\end{equation}
\item if $R_g>D$,
\begin{equation}
W_{rod}= \left\{ 
\begin{array}{ll}
y\frac{L}{b}\left(\frac{D}{b}\right)^{1/3} \, \, \mathrm{for} \, \, y<y_1 \\
y^{9/4}\frac{LD^2}{b^3} +y^{6/4}\frac{DL}{b^2}\, \, \mathrm{for} \, \, y>y_1
\end{array}
\right.
\end{equation}
\end{itemize}
where the volume fraction $y_1\sim\left(\frac{b}{D}\right)^{4/3}$ is defined by the condition $\xi=D$. One can check that the condition $y^*<y_1$ is equivalent to $D<R_g$.

Since the rod (DNA) and flexible polymer (PEG) used in the experiments of reference \cite{alex} satisfy $D\sim R_g $, corresponding to the crossover of the previous expressions of the work of insertion, we can try to approximate the actual work of insertion  by simply superimposing the asymptotic behaviors, following de Vries \cite{devries}.
Therefore the general form of the work of insertion of a cylinder in a polymer solution is given by
\begin{equation}
W=\nu_1 y+\nu_2 y^{6/4}+\nu_3 y^{9/4}
\end{equation}
with the coefficients
\begin{eqnarray}
\label{nu1}\nu_1 & \sim & \frac{LD^2}{Mb^3} \left(1+\frac{R_g}{D}+\left(\frac{R_g}{D}\right)^{5/3}\right)\\
\label{nu2}\nu_2 & \sim & \frac{LD}{b^2}\\
\label{nu3}\nu_3 & \sim & \frac{LD^2}{b^3}
\end{eqnarray}
These  results are identical to those of de Vries, except for our having taken into account explicitly the surface tension of the polymer solution in order to be consistent with the SPT approach.

\section{Numerical estimates}
\label{discussion} 
\begin{figure}
[p!]
\begin{center}
\includegraphics[scale=1]{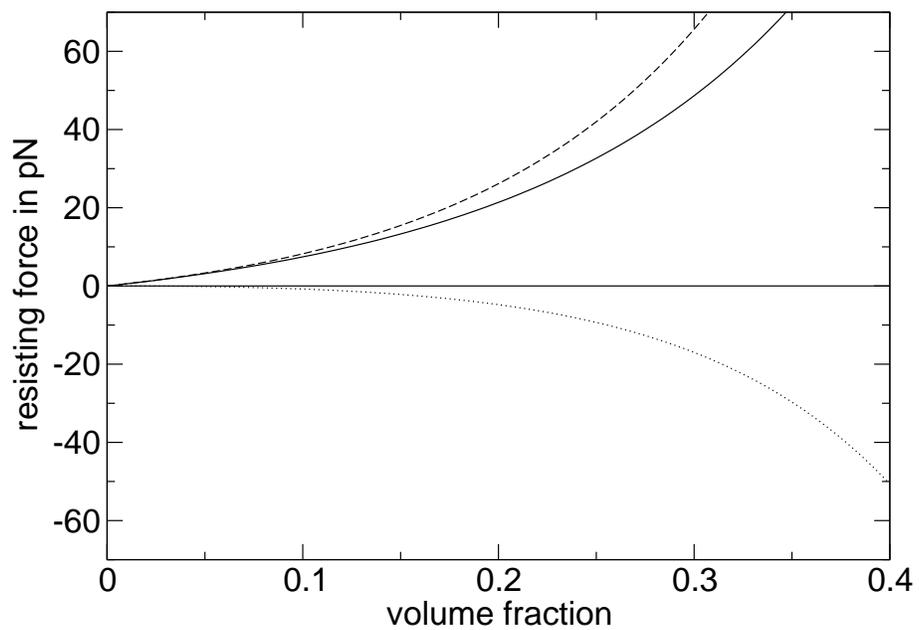}
\caption{Different contributions to the force resisting ejection in $pN$, as a function of volume fraction. The full line is the total force, the long dashed line is the pressure contribution and the short dashed line is the negative surface contribution. $d_{DNA}=2nm\, ,  \, a=1nm$}
\label{volsur}
\end{center}
\end{figure}

\begin{figure}
[p!]
\begin{center}
\includegraphics[scale=1]
{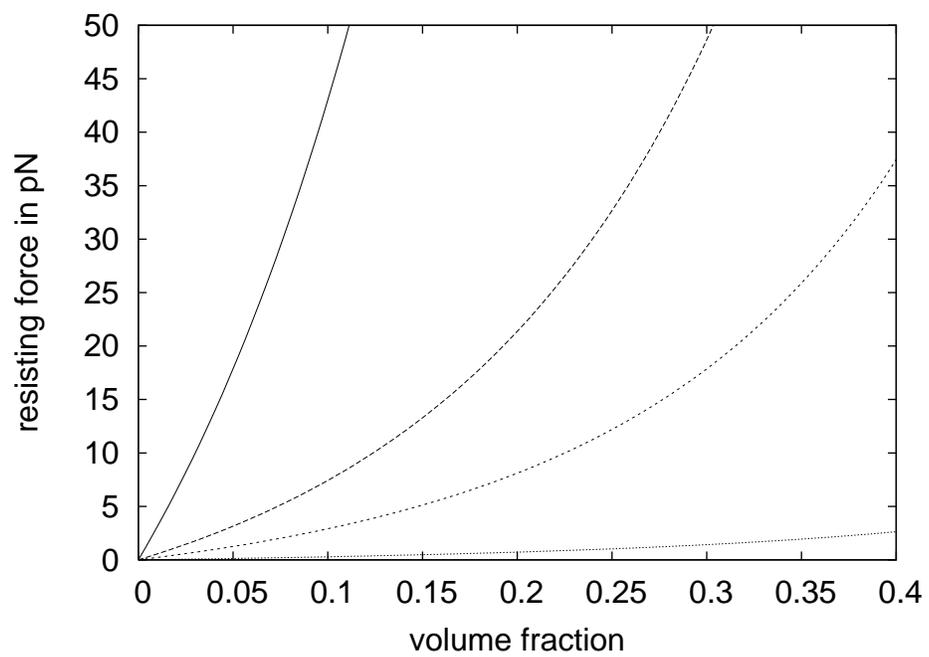}
\caption{Force resisting ejection in $pN$, as a function of volume fraction, for different colloid diameters. From top to bottom $a=0.5nm,a=1nm,a=1.5nm,a=5nm$. $d_{DNA}=2nm$}
\label{fighs}
\end{center}
\end{figure}

We calculated in the last two sections the work of insertion of a cylinder (or a spherocylinder) in a colloidal suspension in both the cases of hard spheres and polymers. The classical interpretation of this quantity is the energetic cost associated with the growth of the particle to its final size and shape in the bulk suspension. This work is therefore associated with a characteristic force that resists the growth of the particle. In the case of the ejection of DNA from a viral capsid into a colloidal suspension, the DNA is ``inserted'' into the suspension along its length. 
Therefore the force associated with the work of insertion is simply given by the derivative of the work with respect to the length of the chain. In the experiments done by Evilevitch \textit{et al.}, the DNA is digested by the enzyme DNase I as it is inserted from the capsid into the colloidal suspension, suggesting that we approximate the ejected DNA as a straight cylinder of diameter $d_{DNA}$. This allows us to neglect the effect of DNA condensation and aggregation. In the case of a hard sphere fluid, the force is calculated by using the SPT expression for the work of insertion of a cylinder:
\begin{eqnarray}
\frac{dW(D,L)}{d(La)}\bigg |_{L\rightarrow \infty} & = & 4 \, \frac{3\left(1+\frac{d_{DNA}}{a}\right)^2(y+y^2+y^3)}{2a(1-y)^3} \nonumber\\
 & & \hspace{-0.5cm} -4 \, \frac{9y^2\left(1+\frac{d_{DNA}}{a}(1+y)\right)}{2a(1-y)^3}\, \, \, \, pN \nonumber\\
 & & \label{forcehs}
\end{eqnarray}
The prefactor 4 gives the force in $pN$, provided that both the diameters of the cylinder $d_{DNA}$ and of the spheres $a$ is expressed in $nm$. Note that a spherocylinder of diameter $d_{DNA}$ in a hard sphere fluid creates a cavity of diameter $d_{DNA}+a$ excluding the center of any hard spheres. The first term in Eq.\ref{forcehs} is the pressure contribution while the second is the surface contribution.
The magnitudes of those two contributions are shown in figure \ref{volsur} for a particular set of diameters. It is clear from this figure that a simple argument where the work of insertion is approximated by a pressure-volume work term overestimates the force, but this effect is not dramatic. 
The influence of the sphere size is shown in figure \ref{fighs}.
The force gets smaller as the sphere size is increased at fixed volume fraction -- fewer particles contribute to the osmotic pressure. If we think of various proteins found \textit{in vivo} behaving purely as hard spheres \cite{ellis}, the range of their sizes from $1$ to $10nm$ shows that for moderate colloid volume fraction (typically 20-30$\%$ \cite{ellis}) the magnitude of the force resisting insertion is of the same order of magnitude as the force driving DNA out of a viral capsid, measured \cite{smith} and estimated \cite{kindt,shelly,phillips} recently to be tens of $pN$ in the first steps of DNA ejection.

\begin{figure}
[p!]
\begin{center}
\includegraphics[scale=1]
{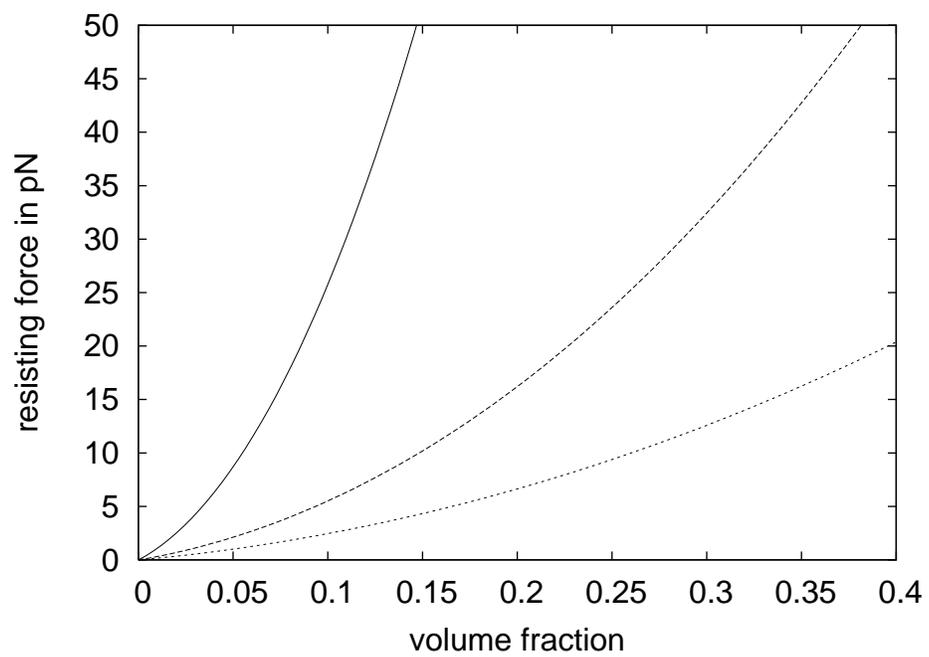}
\caption{Force resisting ejection in $pN$ in a PEG solution, as a function of PEG volume fraction for different monomer diameters. From top to bottom $b=0.2nm,b=0.4nm,b=0.6nm$. $d_{DNA}=2nm$.}
\label{figpol}
\end{center}
\end{figure}

Similarly in the case of polymer solution, the force resisting ejection is given by 
\begin{equation}
f=\nu_1 ' y+\nu_2 ' y^{6/4}+\nu_3 ' y^{9/4}
\end{equation}
with $\nu_i '=\nu_i/L$, the $\nu_i$'s being given by Eqs.\ref{nu1}-\ref{nu3}. Forces $f$ vs volume fraction $y$ is shown in figure \ref{figpol} for different choices of monomer sizes. The value $b=0.4nm$ approximates the statistical size of ethylene glycol monomer \cite{abbott}. As mentioned earlier, the work of insertion of DNA in a solution of PEG has already been calculated by de Vries in his effort to address the polymer-salt-induced DNA condensation problem; his fit of the numerical prefactors to experimental data gives an $f(y)$ close to ours for $b=0.4nm$. Once again, as in the hard sphere case, the force is of order tens of $pN$, comparable to the force ejecting DNA from phage capsids \cite{smith,kindt,shelly,phillips}.
Preliminary experimental results show that DNA ejection from $\lambda$-phage can indeed be inhibited by a PEG solution at moderate concentrations, the force pushing out the DNA being balanced by the force resisting ejection \cite{alex}. At high enough polymer concentration, no ejected DNA is detected. The smallest concentration required for the complete suppression of DNA ejection, up to experimental limits of detection, implies a resisting force between 20 and 30 $pN$. This corresponds to the right order of magnitude of the force driving the DNA out of the capsids when it is fully loaded, as estimated by a careful balance of bending energy and intermolecular repulsions inside the capsid \cite{kindt,shelly,phillips}.

\section{Concluding remarks}
We calculated in this study the work of insertion of a cylinder in a hard sphere fluid and in a polymer solution of arbitrary concentration. In the context of DNA ejection from a viral capsid, this work can be interpreted as the work done against the constant force resisting insertion of the DNA in the suspension. This allows us to evaluate this force by differentiating the work of insertion with respect to chain length. 

In the case of hard spheres, the work of insertion is computed using the SPT, which is known to provide accurate thermodynamical properties of the hard sphere fluid provided that the volume fraction is not close to the close-packing limit. It is shown that the order of magnitude of the work of insertion can be approximated simply through a pressure-volume work term. The force computed in this way is of the order of several $pN$ for sizes of hard spheres ranging from 1$nm$ to 10$nm$. This range of forces compares favorably to the order of magnitude estimated for the ejection force of DNA from bacteriophage.

In the case of flexible polymer solution, the work of insertion is calculated using scaling arguments; the force estimated in this way depends on omitted numerical prefactors which we drop, an omission inherent in the scaling methods. The order of magnitude of the work of insertion can still be evaluated by setting those prefactors equal to unity. We estimate the resisting force exerted by a PEG solution to be of order tens of $pN$ for moderate PEG concentrations. This force is again comparable to the force driving DNA out of capsids \cite{kindt,shelly,phillips}, a result that has been qualitatively confirmed by recent experiments in which it is shown that the amount of ejected DNA decreases with PEG concentration. A more thorough analysis of those experiments will be presented elsewhere using the present theory \cite{thepaper}. 

The spirit of the SPT calculation for the work of insertion in the case of hard spheres and that of the scaling calculation in the case of polymers are very similar. In both cases, the functional form of the work of insertion and its practical evaluation are guided by mathematical and physical requirements. The mathematical requirement in the case of the SPT is the choice of an analytical function to approximate the work, motivated by the appearance of discontinuities only in high order derivatives. In the case of polymers, scaling laws are required because of the critical properties of polymer solutions in the infinite chain length limit. In both cases, the work of insertion matches certain asymptotic and well-understood behaviors that are physically motivated. 

The present calculation provides also a concrete realization of the concept of work of insertion, and of the force associated with it; this concept underlies many modern theories of fluids, but now finds a direct manifestation in the experiments where DNA is ``inserted'' into a colloidal suspension from a viral capsid.

\textbf{Acknowledgments}
Fruitful discussions with A. Ben-Shaul are gratefully acknowledged. This work was partially supported by NSF grant \# CHE9988651 to W.M.G.

\newpage
\section*{Figure Captions}
\begin{itemize}
\item Figure 1: Different contributions to the force resisting ejection in $pN$, as a function of volume fraction. The full line is the total force, the long dashed line is the pressure contribution and the short dashed line is the negative surface contribution. $d_{DNA}=2nm\, ,  \, a=1nm$.
\item Figure 2: Force resisting ejection in $pN$, as a function of volume fraction, for different colloid diameters. From top to bottom $a=0.5nm,a=1nm,a=1.5nm,a=5nm$. $d_{DNA}=2nm$.
\item Figure 3: Force resisting ejection in $pN$ in a PEG solution, as a function of PEG volume fraction for different monomer diameters. From top to bottom $b=0.2nm,b=0.4nm,b=0.6nm$. $d_{DNA}=2nm$.
\end{itemize}
\newpage
\newpage
\newpage
\end{document}